\documentclass[a4paper,aps,pre,floatfix,twocolumn,superscriptaddress,showpacs]{revtex4}

\usepackage[dvips]{epsfig}
\usepackage[dvips]{graphicx}
\usepackage{latexsym,amsmath,verbatim}
\usepackage[dvips,usenames]{color}

\begin{document}

\title{Consensus in networks of mobile communicating agents}

\author{Andrea Baronchelli} 

\affiliation{Departament de F\'isica i Enginyeria Nuclear, Universitat
  Polit\`ecnica de Catalunya, Campus Nord B4, 08034 Barcelona, Spain}
  
  \affiliation{Department of Physics, College of Computer and Information Sciences, Bouv\`e College of Health Sciences, 
Northeastern University, Boston, MA02115, USA}

\author{Albert D\'iaz-Guilera} 

\affiliation{ Departament de F\'isica Fonamental, Universitat de Barcelona, 08028 Barcelona, Spain }

\date{\today}

\begin{abstract}

Populations of mobile and communicating agents describe a vast array of technological and natural systems,
ranging from sensor networks to animal groups. Here, we investigate how a group-level agreement may emerge
in the continuously evolving network deÞned by the local interactions of the moving individuals. We adopt a
general scheme of motion in two dimensions and we let the individuals interact through the minimal naming
game, a prototypical scheme to investigate social consensus. We distinguish different regimes of convergence
determined by the emission range of the agents and by their mobility, and we identify the corresponding scaling
behaviors of the consensus time. In the same way, we rationalize also the behavior of the maximum memory
used during the convergence process, which determines the minimum cognitive/storage capacity needed by the
individuals. Overall, we believe that the simple and general model presented in this paper can represent a helpful
reference for a better understanding of the behavior of populations of mobile agents.

\end{abstract}

\pacs{89.75.-k, 05.65.+b, 89.65.-s, 89.75.Hc}

\maketitle

\section{Introduction}

Autonomous mobile and communicating agents provide extremely efficient solutions to a wide range of technological problems by guaranteeing robustness, flexibility, and dynamic adaptability \cite{lange1999seven}.  A typical case is that of a population of  robots that have to explore an unknown environment, and cope with situations that by definition can not be foreseen in advance \cite{steels2001language,steels2003evolving}. For example, robots could have to negotiate a common lexicon to name different places of the environment they are surveying, and then use this shared linguistic knowledge to carry out goal-directed behavior \cite{lingodroids2011,schulz2011lingodroids}. In the same way, also the performances of sensor networks  \cite{akyildiz2002survey} can be enhanced by the introduction of mobile agents \cite{tong2003sensor,leonard2007collective}. In natural systems, on the other hand, mobile populations that coordinate through chemical or audible signals are obviously widespread, ranging from cell populations \cite{deisboeck2009collective} to animal groups \cite{giardina2008collective}.

In all of these cases, mobile agents locally broadcast their signal to nearby nodes \cite{lim2007preferential,lu2008naming,baronchelli2011role}. Thus, communication takes place on a continuously evolving network  whose properties are determined by such parameters as the emission range of the individuals, their mobility, or their density. Network theory \cite{caldarelli2007sfn,barratbook,mendesbook,romuvespibook,boccaletti2006cns} is therefore the natural framework to investigate the emerging population-scale properties of the system.  However, previous research has so far focused mainly on static random geometric networks \cite{lim2007preferential,lu2008naming}, or on the opposite case of rapidly changing structures \cite{belykh2004connection,frasca2008synchronization}. Only very recently the more general case of time-dependent networks has been fully addressed, for the specific case of the synchronization of mobile oscillators \cite{albert2011sync}.

In this paper we address the fundamental problem of social consensus. To this purpose, we model individuals that move in a 2-dimensional plane and have to agree on a given convention by performing standard language games \cite{wittgenstein53english,Steels1996}. For example, they might be in need of creating or selecting autonomously a key for encrypted communication  \cite{lu2008naming}, or to independently elect a leader \cite{angluin1980global}. We then study how different parameters of the mobile agents, in particular their velocity and their communication range, affect the overall agreement process. We are able to identify different regimes ruling the consensus dynamics, and we rationalize our findings by considering the properties of the communication networks. 

\section{The model}

We model the mobility of the agents according to the general scheme put forth in \cite{albert2011sync}. A population of $N$ individuals move in a 2-dimensional $L \times L $ box with periodic boundary conditions. The velocity of each agent is $v$. The angle of the $i$th agent's motion is $\xi_i(t_k) \in [0,2\pi]$, and it changes randomly at discrete time steps $t_{k}$ 
($t_{k+1}-t_k=\tau_M$). Thus, the evolution in time of the i$th$ agent's position is:

\begin{eqnarray}
 x_i(t_k+\Delta t) &=& x_i(t_k) + v \cos \xi_i(t_k) \Delta t \;\;\; \mbox{mod } L \\
y_i(t_k+\Delta t) &=& y_i(t_k) + v \sin \xi_i(t_k) \Delta t \;\;\;  \mbox{mod } L,
\label{e:move}
\end{eqnarray}

\noindent where $\Delta t \le \tau_M$. The motion of the individuals is therefore diffusive, the diffusion coefficient being $D \sim v^2 \tau_M$.

\begin{figure}[!t]
\begin{center}
\includegraphics*[width=0.42\textwidth]{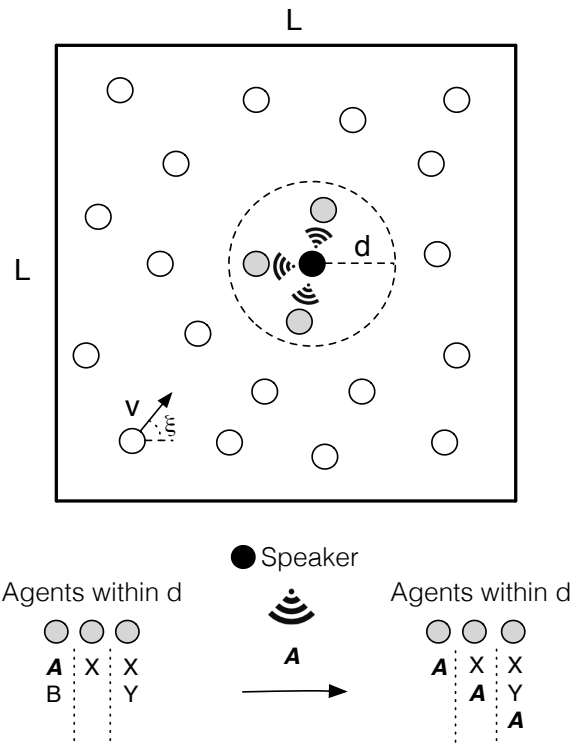}
\end{center}
\caption{The model.  $N$ agents (circles) move with velocity $v$ and randomly assigned angles $\xi_i$ in a box of size $L$, with periodic boundary conditions. At each time step one of the agents is chosen as a speaker (black circle) and emits a word randomly extracted from her inventory ({\it \textbf{A}}, in figure). All the agents within a distance $d$ (gray circles) receive the word, and update their inventories as shown in the schematic representation below the box. If an agent already knows that word, she deletes all the competing synonyms in her inventory, otherwise she simply adds the new word to it. No feedback is provided to the speaker, whose inventory is not altered. }
 \label{f:model}
 \end{figure}

The agents play the minimal naming game (NG) \cite{Baronchelli_JStatMech_2006,Baronchelli_ng_long} implemented with local broadcasting without feedback \cite{baronchelli2011role}. Each agent is characterized by an inventory of words (or ``conventions", ``opinions'', ``forms" or ``states"). At the beginning all inventories are empty. At discrete time steps of duration $\tau_S$ an agent is randomly selected as speaker. She selects randomly a word from her inventory, and transmits it to all the agents within a distance $d$ from her position (if the inventory is empty she invents a brand new word and stores it into her own inventory, before broadcasting it). Each receiving agent updates her state depending on whether her inventory contains or does not contain the transmitted word. In the first case, the agent deletes all the competing synonyms and keep only that word into her inventory. In the latter case, on the other hand, she adds the new received word to her inventory. The speaker receives no feedback about her emission, and consequently does not modify her inventory Figure \ref{f:model} summarizes the rules of the model.

For simplicity, in this paper we choose  $\Delta t = \tau_M =  \tau_S =1$. This means that in a time step (i) all agents move in straight line, (ii) all agents are reassigned a random angle, and (iii) \textit{one} agent broadcasts to her neighbors. 
 
\section{Paths to consensus}

\begin{figure}[!t]
\begin{center}
\includegraphics*[width=0.42\textwidth]{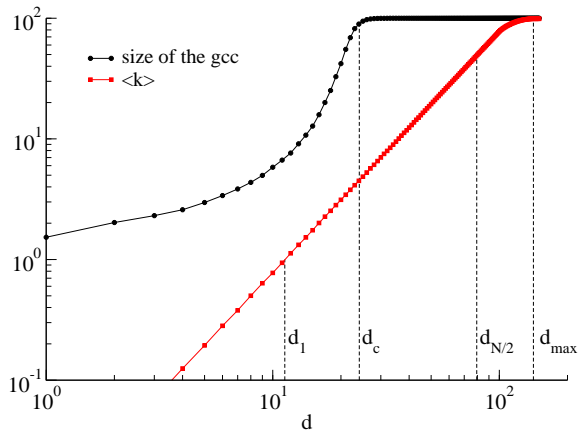}
\end{center}
\caption{(Color online) Properties of the static network.  At $d>d_1$ every emission is heard on average by at least an individual of the population, while at $d=d_{N/2}$ a majority of agents listens to each communication act. At $d=d_c$ the giant connected component (``gcc") is formed by $N$ nodes (in reality, due to finite size effects, this happens for a slightly larger emission range). At $d_{max}$  the network is fully connected.}
 \label{f:distances}
 
 \end{figure}

The NG is an ordering process. An initially disordered configuration ends up in a consensus, ordered, state in which everybody has the same unique word \cite{Baronchelli_JStatMech_2006,de2006reach}. The consensus (or ``convergence") time, $t_{conv}$, is therefore a crucial quantity. Also important is the maximum number of words agents have to store, $M$, which describes the global amount of memory needed by the system to reach a consensus. Previous studies on quenched graphs have shown that both quantities depend dramatically on the topology of the social network describing the possible interactions between the individuals \cite{dallasta06b}. Consequently, to investigate the properties of a mobile population it is convenient to focus on the properties of the static network describing the instantaneous communications of the agents. This is the graph that is obtained, at any time, by drawing an undirected link between any two agents that are closer than the emission range $d$.  Recalling that the average degree (i.e., the average number of neighbors of a randomly selected node) of the network is simply $\langle k \rangle = \pi N d^2 / L^2$, these values of $d$ identify different scenarios

\begin{itemize}
\item $d_1 \equiv d_{\langle k \rangle=1}$ is the range above which the average degree is larger than $1$, so that every emission is received on average by some agent.
\item $d_c \equiv d_{\langle k \rangle \simeq 4.51}$ is the critical value for a percolation transition, yielding a giant component of size $N$ \cite{dall2002random}. 
\item $d_{N/2} \equiv d_{\langle k \rangle=N/2}$ is the point where every communication involves on average the majority of the population. 
\item $d_{max} \equiv d_{\langle k \rangle=N}$ is the value that yields a fully connected network. It holds $d_{max}=\sqrt{L^2/2}$.
\end{itemize}

\noindent Of course, it holds $d_1<d_c<d_{N/2}<d_{max}$. In this paper we fix $L=200$ and $N=100$, unless where explicitly stated, so that $d_1 \simeq 11.3$, $d_c \simeq 24.1$, $d_{N/2} \simeq 79.8$ and $d_{max} \simeq 141.4$.  Figure \ref{f:distances} shows the dependence of the average degree and the size of the giant component on the parameter $d$ for this choice of $L$ and $N$. 

For a qualitative partitioning of the observed phenomenology in terms of distinct regimes, it is convenient to consider two timescales. One describes the stability of a cluster of agents, and the other accounts for the time over which a consensus is reached within the same cluster \footnote{Notice that the time is normalized so that every NG interaction is alternated with a diffusion step, due to the choice $\Delta t = \tau_M =  \tau_S =1$.}. Their ratio $\eta$ assesses therefore the impact of local, intra-cluster, activity on global, inter-cluster,  dynamics. For the robustness of a cluster, we consider the average number of timesteps needed by an individual to leave a group of size $n(d)$ (an increasing function of $d$ for $d<d_c$), that scales as $t_{1} \sim n(d) / v^2$ \cite{albert2011sync}.
For the within-cluster average consensus time, on the other hand, we note that it can be treated as independent from the cluster size $n(d)$, $t_{conv} \sim const.$, both when the considered groups are densely connected and when they are very small. The reason is that in the first case the broadcasting rule brings about a very fast consensus time, which becomes instantaneous in fully connected graphs \cite{baronchelli2011role}, while in the latter case consensus is quick simply because just a few agents have to agree \cite{Baronchelli_JStatMech_2006,baronchelli2011role}. This approximation is appropriate for our purposes since we aim at defining a qualitative index able to discriminate between extreme regimes. Moreover, as we shall see, it is further validated by the results discussed in the following of the paper (Sec. \ref{sec:size}).   Therefore, for the ratio $\eta$ between the two timescales it holds:

\begin{equation}
 \eta = \frac{t_{1}}{t_{conv}} \sim \frac{n(d)}{v^2},
 \label{e:eta}
\end{equation}

\noindent which is obviously an increasing function of $d$ as $d<d_c$.

\begin{figure}[!t]
\begin{center}
\includegraphics*[width=0.45\textwidth]{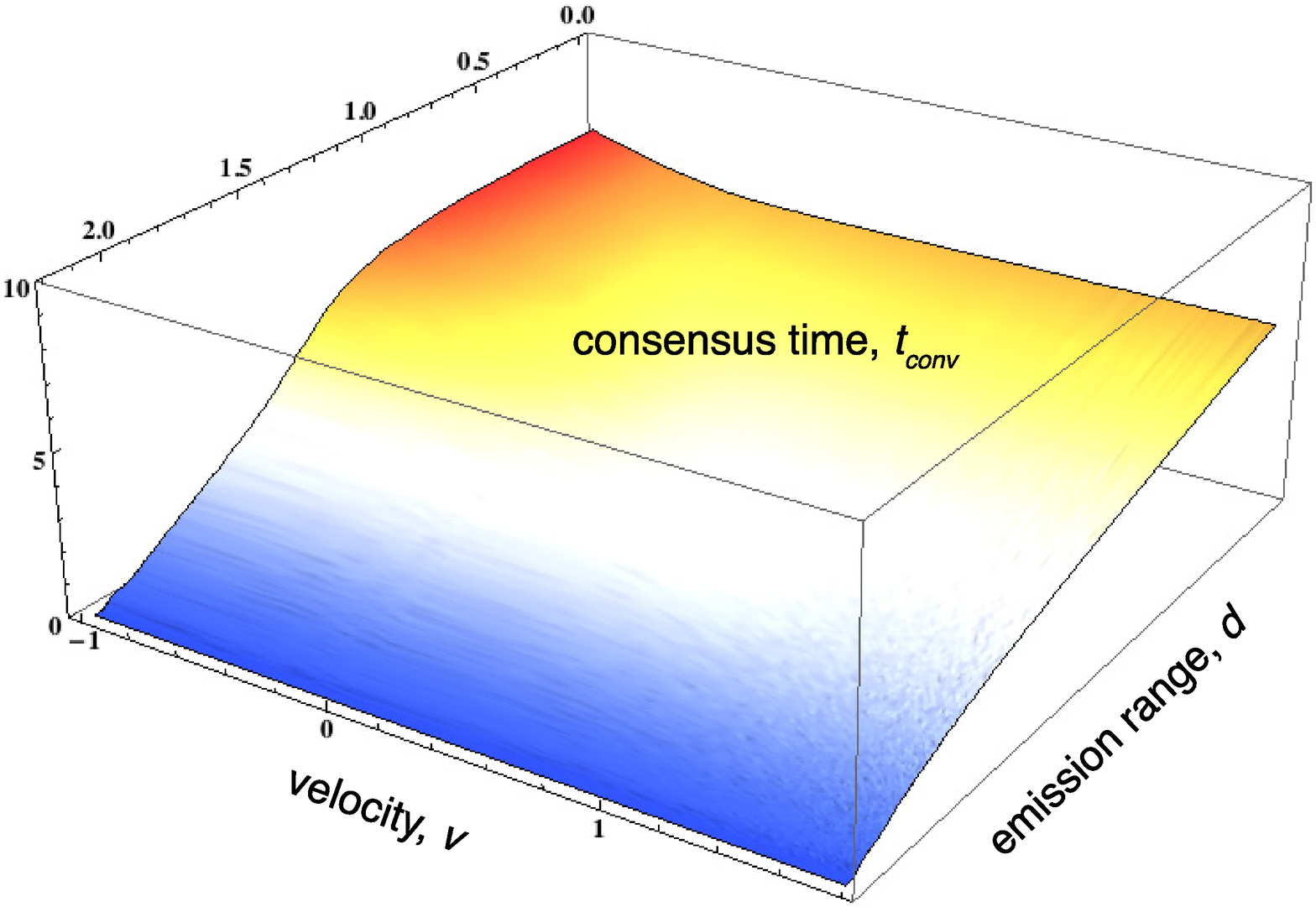} 
\includegraphics*[width=0.45\textwidth]{fig3b.eps}
\end{center}
\caption{(Color online) Top: Consensus time as a function of $d$ and $v$ (all axes report the $\log_{10}$ of the respective quantities). Bottom: $t_{conv}$ as a function of $d$ for different for different values of the agents velocity $v$. Dotted vertical lines represent $d_1$, $d_c$ and $d_{N/2}$.}
\label{f:consensus_raw}
\end{figure}

\subsection{Consensus time}

 \begin{figure}[!t]
\begin{center}
\includegraphics*[width=0.45\textwidth]{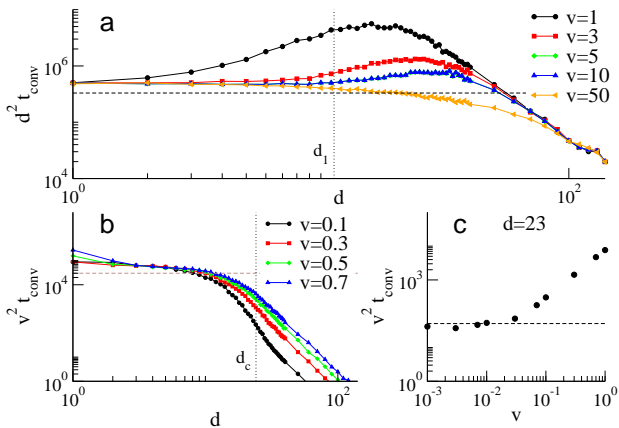}
\end{center}
\caption{(Color online) Rescaling of the consensus time, $d<d_c$. When $\eta \ll 1$,  the $d^2 t_{conv} \sim const.$ and curves for large velocities collapse (a). For $\eta>1$, on the other hand, curves for different, and small values, of $v$ collapse as the consensus time is rescaled as $v^2 t_{conv} \sim const$ (b). This behavior is observed also for values of $d$ close to $d_c$ provided that small enough velocities are considered (c). In all panels horizontal dashed lines represent a constant behavior, and serve as a guide for the eye.}
 \label{f:low_d}
 \end{figure}

Figure \ref{f:consensus_raw} shows the behavior of the consensus time $t_{conv}$ as a function of the emission range $d$, and for different values of the agents' velocity $v$. The consensus is fast for large values of $d$ (and becomes instantaneous as soon as $d = d_{max}$, when everybody receives the word transmitted by the first speaker), but it increases for shorter ranges, in a way that crucially depends upon the $v$ parameter. We can identify three regimes.

(1) $\eta \ll 1$ holds for small $d$ and large $v$, and implies a rapidly evolving network. Agents continuously change their neighbors, and hence the partners of their communication acts, pretty much as in the case of an annealed network. Thus, consensus emerges through \textit{global agreement} at the system size level, after the agents have correlated their inventories so as to allow for successful communication to take place \cite{dallasta06c}. As $d<d_1$ the behavior $t_{conv} \sim 1/\langle k \rangle = 1/d^2$ is observed (Figure \ref{f:low_d}, a), describing the existence of empty communication acts (unheard emissions) when on average each node has less than one neighbor.

(2) $\eta > 1$  and $d < d_c$, on the other hand, implies smaller velocities. In this case, small and \textit{isolated clusters} of agents locally reach an agreement on different conventions. Global consensus emerges at a later time through the competition between these words, in a situation reminiscent of what happens in low-dimensional lattices \cite{ng_lowdim}. The intra-cluster movements determine the leading timescale, implying a scaling of the form $t_{conv} \sim 1/v^2$ (Figure \ref{f:low_d}, b and c).

\begin{figure}[!t]
\begin{center}
\includegraphics*[width=0.45\textwidth]{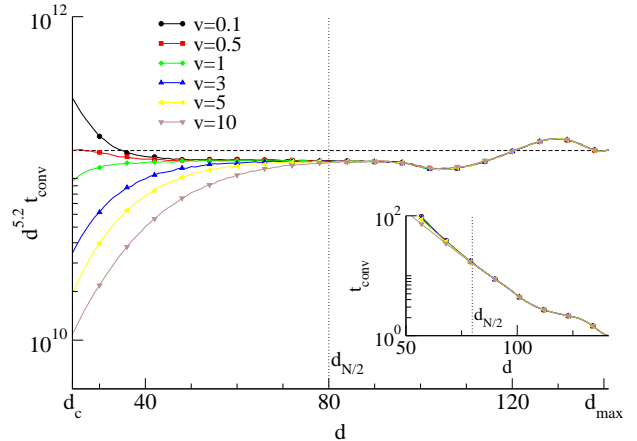}
\end{center}
\caption{(Color online) Rescaling of the consensus time, $d>d_c$. For large $\eta >1$ and large $d$ curves for different values of $v$  behave as $t_{conv} \sim 1/d^{5.2}$ as far as $d$ (and hence $\langle k \rangle$) is not too large. The behavior is better observed  for smaller velocities, since large values of $v$ reduce the value of $\eta$. As $d>d_{N/2}$ the different curves collapse, and $v$ becomes an irrelevant parameter, as shown also in the inset with non-rescaled abscissas. Horizontal dashed lines represent a constant behavior, and serve as a guide for the eye.}
 \label{f:large_d}
 \end{figure}

(3) $\eta > 1$  and $d \gg d_c$, finally, identify a scenario in which the whole population forms a \textit{single connected cluster}, describing a random geometric graph. In \cite{lu2008naming}  Lu and coworkers showed numerically that $t_{conv} \sim 1/ \langle k \rangle^{2.6}$ (``when $\langle k \rangle \ll N$") for static random geometric graphs \footnote{The broadcasting rule of \cite{lu2008naming} is slightly different from the one presented in \cite{baronchelli2011role} and implemented here. The results presented here indicate that the difference is irrelevant in this context.}. Accordingly, in Figure \ref{f:large_d} we observe the behavior $t_{conv} \sim 1/d^{5.2}$, which as expected degrades before $d_{max}$, where $\langle k \rangle = N-1$. It may be further noted that, as $d>d_{N/2}$, the curves for different $v$ behave identically (Figure \ref{f:large_d}, inset). This is due to the fact that here the first speaker transmits her word to an absolute majority of the agents, which on their turn drive the system to consensus very rapidly thanks to  the fact that the NG is a drift-driven process \cite{Baronchelli_ng_long}.

\subsection{Memory usage}

\begin{figure}[!t]
\begin{center}
\includegraphics*[width=0.45\textwidth]{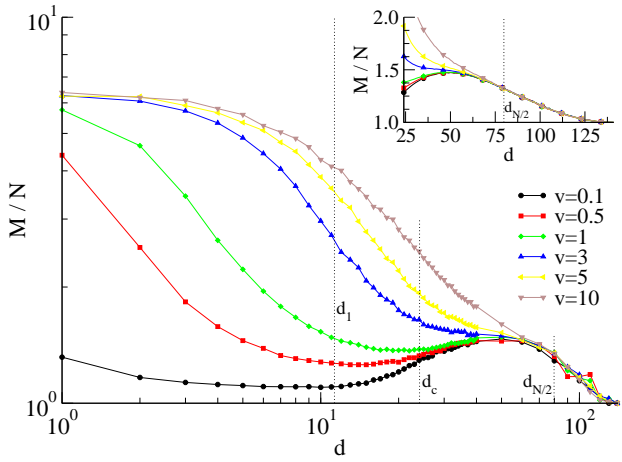}
\end{center}
\caption{(Color online) Memory usage. The average maximum memory used by each agent is plotted for different values of the velocity $v$. While for large mobility the individual storage capacity monotonously increases as $d$ is reduced, small velocity induces a more complex behavior. In this case, the curves exhibit a minimum for $d \simeq d_1$ and a maximum for $d \simeq 60$, for the usual choice of the parameters. For $d>d_{N/2}$ curves for different velocities collapse (see Inset). }
 \label{f:memory}
 \end{figure}

The agents get to know different words at the same time during the process that eventually leads them to a consensus, and the NG rules do not fix a limit to the size to their inventory. Therefore, it is important to look at the maximum memory consumption that the population experiences during the whole process, corresponding to the maximum number of words, $M$, present in the system, i.e., to the sum of the inventory sizes of the $N$ agents. Gaining quantitative insights into the behavior of this quantity is more difficult than for $t_{conv}$, but some hints can be gained from numerical investigations. Figure \ref{f:memory} shows the average maximum memory per agent as a function of $d$, and for different values of $v$. 

Again, to understand what goes on in the case of mobile agents it is helpful to recall the results obtained in static networks. It turns out that a finite connectivity implies finite memory requirements \cite{dallasta06b}, while a fully connected graph would require infinite inventories (in the thermodynamic limit) \cite{Baronchelli_JStatMech_2006}. In general, a larger average degree requires a bigger memory effort for the agents  \cite{dallasta06b}. Of course, however, the broadcasting rule implies an immediate consensus on fully connected graphs, entailing a minimal amount of memory. Moreover, ordered low-dimensional lattices determine an extremely reduced use of memory, since convergence is reached through the competition of different clusters of agents who have reached a local consensus and therefore store one word only \cite{ng_lowdim,dallasta06c}.

In the light of these results, it is possible to rationalize the findings presented in Figure \ref{f:memory}. As expected, when the dominant process is \textit{global agreement} ($\eta \ll 1$, small $d$ and large $v$), more memory is needed [region (1), above]. Every agent is exposed to a large number of words due to the high mobility of the population. As the velocity is decreased, and the system enters the phase of \textit{isolated clusters}, on the other hand, consensus is reached with a smaller memory demand, due to the early onset of regions of local agreement [region (2)]. As the emission range is increased, and one \textit{single connected cluster} emerges, finally, curves for different values of $v$ become more and more similar, and collapse as $d>d_{N/2}$ (Figure \ref{f:memory}, inset), as observed for $t_{conv}$ [region (3)]. Curves for small $v$ exhibit a peak  somewhere in the region is $d_c < d <d_{N/2}$ (more precisely for $50 < d < 60$ in Figure \ref{f:memory}), separating the memory-favorable cases of a fully connected graph and a collection of isolated clusters. However, as $d$ is further decreased, the memory requirements grow again. Qualitatively, this behavior can be understood considering that when $d<d_1$, local clusters are in general tree-like, so that more words can coexist in each of them before a local agreement emerges.

\begin{figure}[!t]
\begin{center}
\includegraphics*[width=0.45\textwidth]{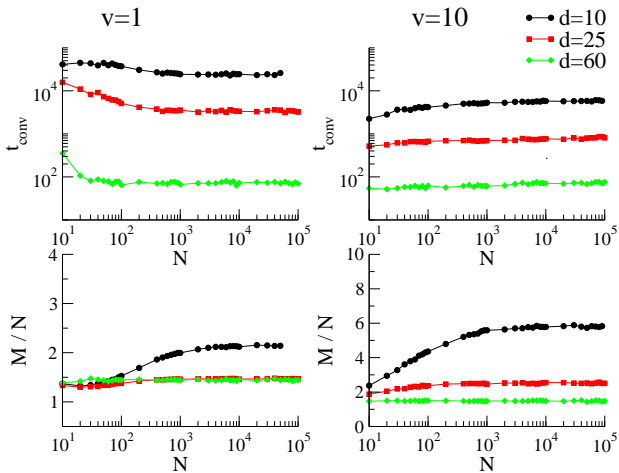}
\end{center}
\caption{(Color online) Role of the population size (at fixed $L$). Top: The consensus time reaches a plateau at large $N$ in the case of small velocities (left panel), while it grows very weakly for larger mobilities (right panel). Bottom: The maximum memory per agent required during the process is constant at large $N$ for both high and low mobility rates. }
 \label{f:popsize}
 \end{figure}
 
\subsection{Role of the population density}
\label{sec:size}

In finite static networks the global consensus is reached only if $d > d_c$, when all the nodes belong to the same unique giant connected component.  Previous studies \cite{lu2008naming}, conducted adopting a slightly modified NG protocol, showed a $N^{\alpha}$, with $\alpha \simeq 2.1$, dependence of the consensus time on the population size $N$ in random geometric graphs with fixed average connectivity (and consequently varying $L$), and, as mentioned above, an important role also for the average degree, $t_{conv} \sim \langle k \rangle^{-2.6}$ (as far as $k \ll N$) for fixed $L$. The scaling of $t_{conv}$ with the population size agrees well with a theoretical argument put forth in \cite{lu2008naming}, which is in its turn very similar to the analysis of the NG in low-dimensional lattices \cite{ng_lowdim}. The crucial point is that different clusters of local consensus emerge rapidly, and the global agreement is the result of a cluster-cluster competition. The presence of mobility makes this whole argument break down, and the increased population mixing yields a faster scaling of the consensus (data not shown), in agreement to what is observed in small-world networks \cite{dallasta06,lu2008naming}.  

Crucially, on the other hand, mobility guarantees that the consensus is always reached, at least asymptotically, independently from the emission range of the individuals. However, it is important noting that population size and average degree are intimately connected in the framework we are addressing, where the box side $L$ is kept constant. Indeed, as discussed above, it holds $\langle k \rangle \propto N$, and therefore 

\begin{equation}
d_1 \sim d_c \sim \frac{1}{\sqrt{N}} \stackrel{N \rightarrow \infty}{\longrightarrow} 0,
\end{equation}

\noindent and $d_{N/2}$ being the only characteristic length surviving in the thermodynamic limit. The ``small" range regimes we have described would therefore vanish for very large population size. 

Concretely, the case of very large population densities is quite unrealistic from the point of view of any application involving mobile individuals. However, for the sake of completeness, we report in Figure \ref{f:popsize} the results of numerical investigations for different values of $d$ and $v$. For small velocities (left column), very small population sizes are not efficient since the individuals waste time in finding each other. As $d$ grows, however, $t_{conv}$ saturates. After a certain threshold, the broadcasting rule makes the actual number of agents present in the population irrelevant. A similar behavior is observed for the case of a larger mobility of the individuals (right column). Here, however, smaller population sizes converge slightly faster since agents get more easily in contact, and the dynamic of the NG favors a smaller number of competing conventions. In addition, the augmented population mixing is responsible for the fact that, (i) the consensus time does not reach a plateau but keeps increasing very slowly with the system size, as $t_{conv} \sim N^{\alpha}$, with $\alpha < 0.05$ and (ii) the memory consumption is larger, in agreement with what is observed in the case of static networks for the NG with pairwise interactions \cite{dallasta06b}. Finally, it is worth noting that the very weak dependence of the consensus time on $N$ validates, \textit{a posteriori}, the assumption on the constant behavior of $t_{conv}$ that yielded to Eq.~\ref{e:eta}.


%

\section{Conclusion}

In this paper we have analyzed the consensus problem in a population of autonomous mobile agents. We have focused on the crucial case of a self-organized agreement process, to be established without any central control nor coordination. Agents move with the same velocity and different, randomly changing, angles in a two dimensional space, and communicate through the NG protocol, locally broadcasting in an circle of radius $d$. We have shown that different characteristic emission ranges exist, defining, together with the mobility rates of the agents, the boundaries between different consensus regimes. In particular, we have highlighted three main mechanisms ruling the onset of consensus. If the emission range is small, a rapidly mixing population will undergo a global agreement process, while a slower mobility will bring about the appearance of isolated clusters in which a local consensus on different conventions forms rapidly, the final consensus resulting from the competitions of these clusters. Finally, large emission ranges establish a single connected cluster where the static limit of random geometric networks is recovered. Accordingly, we have pointed out the scaling relations of the consensus time in each region and rationalized the memory needs of the agents. Finally we have considered the role of the population density, showing that, due to the broadcasting rule, even unrealistically high densities have tiny impact on both the consensus time and the memory consumption of the agents. 

Examples of populations of mobile and communicating agents pop out in many natural contexts, chiefly in cases of group of animals or micro-organisms. Yet, it is perhaps the technological advancement in the fields of robotics and telecommunications that makes the investigation of this issue increasingly urgent. From this point of view, the results presented in this paper may be far-reaching. Both the mobility and the communication models we have adopted, indeed, are straightforward and might serve as a reference to gain insights into more complex and realistic models. At the same time, the scheme we have introduced can be extended in the future so as to address such issues as the coupling between motion and communication. For example, our results show that the larger the agents' mobility the shorter the consensus time, so that  changing neighbors frequently turns out to be an efficient strategy. Thus, agents with a tendency to move apart from each other after a success 
could expedite convergence, preventing the formation of local clusters. On the contrary, individuals with a tendency to reinforce existing links by aligning their direction of motion after a successful interaction would probably lead to the emergence of different swarms or flocks, internally agreeing on different conventions, that would move apart from each other and hinder the onset of a global agreement. Along the same lines, finally, another possibility for future work could imply the adoption and study of higher level communication protocols, such as spatially oriented language games \cite{lingodroids2011,schulz2011lingodroids}.

 \textit{Acknowledgments.} A. Baronchelli acknowledges support from the Spanish Ministerio de
  Ciencia e Innovaci\'{o}n through the Juan de la Cierva program, as
  well as from project  FIS2010-21781-C02-01 (Fondo Europeo de
  Desarrollo Regional) and from the Junta de Andaluc\'{i}a project
  P09-FQM4682. A. D\'iaz-Guilera acknowledges support from the Spanish DGICyT Grant  FIS2009-13730, and from the Generalitat de Catalunya 2009SGR00838.



\begin{thebibliography}{31}
\expandafter\ifx\csname natexlab\endcsname\relax\def\natexlab#1{#1}\fi
\expandafter\ifx\csname bibnamefont\endcsname\relax
  \def\bibnamefont#1{#1}\fi
\expandafter\ifx\csname bibfnamefont\endcsname\relax
  \def\bibfnamefont#1{#1}\fi
\expandafter\ifx\csname citenamefont\endcsname\relax
  \def\citenamefont#1{#1}\fi
\expandafter\ifx\csname url\endcsname\relax
  \def\url#1{\texttt{#1}}\fi
\expandafter\ifx\csname urlprefix\endcsname\relax\def\urlprefix{URL }\fi
\providecommand{\bibinfo}[2]{#2}
\providecommand{\eprint}[2][]{\url{#2}}

\bibitem[{\citenamefont{Lange and Oshima}(1999)}]{lange1999seven}
\bibinfo{author}{\bibfnamefont{D.}~\bibnamefont{Lange}} \bibnamefont{and}
  \bibinfo{author}{\bibfnamefont{M.}~\bibnamefont{Oshima}},
  \bibinfo{journal}{Comm. of the ACM} \textbf{\bibinfo{volume}{42}},
  \bibinfo{pages}{88} (\bibinfo{year}{1999}).

\bibitem[{\citenamefont{Steels}(2001)}]{steels2001language}
\bibinfo{author}{\bibfnamefont{L.}~\bibnamefont{Steels}},
  \bibinfo{journal}{Intelligent Systems, IEEE} \textbf{\bibinfo{volume}{16}},
  \bibinfo{pages}{16} (\bibinfo{year}{2001}).

\bibitem[{\citenamefont{Steels}(2003)}]{steels2003evolving}
\bibinfo{author}{\bibfnamefont{L.}~\bibnamefont{Steels}},
  \bibinfo{journal}{Trends in Cog. Sci.} \textbf{\bibinfo{volume}{7}},
  \bibinfo{pages}{308} (\bibinfo{year}{2003}).

\bibitem[{\citenamefont{Schulz et~al.}(2011)\citenamefont{Schulz, Glover,
  Milford, Wyeth, and Wiles}}]{lingodroids2011}
\bibinfo{author}{\bibfnamefont{R.}~\bibnamefont{Schulz}},
  \bibinfo{author}{\bibfnamefont{A.}~\bibnamefont{Glover}},
  \bibinfo{author}{\bibfnamefont{M.}~\bibnamefont{Milford}},
  \bibinfo{author}{\bibfnamefont{G.}~\bibnamefont{Wyeth}}, \bibnamefont{and}
  \bibinfo{author}{\bibfnamefont{J.}~\bibnamefont{Wiles}}, in
  \emph{\bibinfo{booktitle}{ICRA 2011, The International Conference on Robotics
  and Automation, Shanghai, China}} (\bibinfo{year}{2011}), pp.
  \bibinfo{pages}{178--183}.

\bibitem[{\citenamefont{Schulz et~al.}(2011{\natexlab{b}})\citenamefont{Schulz,
  Wyeth, and Wiles}}]{schulz2011lingodroids}
\bibinfo{author}{\bibfnamefont{R.}~\bibnamefont{Schulz}},
  \bibinfo{author}{\bibfnamefont{G.}~\bibnamefont{Wyeth}}, \bibnamefont{and}
  \bibinfo{author}{\bibfnamefont{J.}~\bibnamefont{Wiles}},
  \bibinfo{journal}{Adaptive Behavior}   \textbf{\bibinfo{volume}{19}},
  \bibinfo{pages}{409}(\bibinfo{year}{2011}{\natexlab{b}}).
  
\bibitem[{\citenamefont{Akyildiz et~al.}(2002)\citenamefont{Akyildiz, Su,
  Sankarasubramaniam, and Cayirci}}]{akyildiz2002survey}
\bibinfo{author}{\bibfnamefont{I.}~\bibnamefont{Akyildiz}},
  \bibinfo{author}{\bibfnamefont{W.}~\bibnamefont{Su}},
  \bibinfo{author}{\bibfnamefont{Y.}~\bibnamefont{Sankarasubramaniam}},
  \bibnamefont{and} \bibinfo{author}{\bibfnamefont{E.}~\bibnamefont{Cayirci}},
  \bibinfo{journal}{IEEE Comm. Mag.}
  \textbf{\bibinfo{volume}{40}}, \bibinfo{pages}{102} (\bibinfo{year}{2002}).

\bibitem[{\citenamefont{Tong et~al.}(2003)\citenamefont{Tong, Zhao, and
  Adireddy}}]{tong2003sensor}
\bibinfo{author}{\bibfnamefont{L.}~\bibnamefont{Tong}},
  \bibinfo{author}{\bibfnamefont{Q.}~\bibnamefont{Zhao}}, \bibnamefont{and}
  \bibinfo{author}{\bibfnamefont{S.}~\bibnamefont{Adireddy}}, in
  \emph{\bibinfo{booktitle}{Military Communications Conference, 2003. MILCOM
  2003. IEEE}} (\bibinfo{organization}{IEEE}, \bibinfo{year}{2003}),
  vol.~\bibinfo{volume}{1}, pp. \bibinfo{pages}{688--693}.

\bibitem[{\citenamefont{Leonard et~al.}(2007)\citenamefont{Leonard, Paley,
  Lekien, Sepulchre, Fratantoni, and Davis}}]{leonard2007collective}
\bibinfo{author}{\bibfnamefont{N.}~\bibnamefont{Leonard}},
  \bibinfo{author}{\bibfnamefont{D.}~\bibnamefont{Paley}},
  \bibinfo{author}{\bibfnamefont{F.}~\bibnamefont{Lekien}},
  \bibinfo{author}{\bibfnamefont{R.}~\bibnamefont{Sepulchre}},
  \bibinfo{author}{\bibfnamefont{D.}~\bibnamefont{Fratantoni}},
  \bibnamefont{and} \bibinfo{author}{\bibfnamefont{R.}~\bibnamefont{Davis}},
  \bibinfo{journal}{Proceedings of the IEEE} \textbf{\bibinfo{volume}{95}},
  \bibinfo{pages}{48} (\bibinfo{year}{2007}).

\bibitem[{\citenamefont{Deisboeck and Couzin}(2009)}]{deisboeck2009collective}
\bibinfo{author}{\bibfnamefont{T.}~\bibnamefont{Deisboeck}} \bibnamefont{and}
  \bibinfo{author}{\bibfnamefont{I.}~\bibnamefont{Couzin}},
  \bibinfo{journal}{Bioessays} \textbf{\bibinfo{volume}{31}},
  \bibinfo{pages}{190} (\bibinfo{year}{2009}).

\bibitem[{\citenamefont{Giardina}(2008)}]{giardina2008collective}
\bibinfo{author}{\bibfnamefont{I.}~\bibnamefont{Giardina}},
  \bibinfo{journal}{HFSP Journal} \textbf{\bibinfo{volume}{2}},
  \bibinfo{pages}{205} (\bibinfo{year}{2008}).

\bibitem[{\citenamefont{Lim et~al.}(2007)\citenamefont{Lim, Braha, Wijesinghe,
  Tucker, and Bar-Yam}}]{lim2007preferential}
\bibinfo{author}{\bibfnamefont{M.}~\bibnamefont{Lim}},
  \bibinfo{author}{\bibfnamefont{D.}~\bibnamefont{Braha}},
  \bibinfo{author}{\bibfnamefont{S.}~\bibnamefont{Wijesinghe}},
  \bibinfo{author}{\bibfnamefont{S.}~\bibnamefont{Tucker}}, \bibnamefont{and}
  \bibinfo{author}{\bibfnamefont{Y.}~\bibnamefont{Bar-Yam}},
  \bibinfo{journal}{Europhys. Lett.} \textbf{\bibinfo{volume}{79}},
  \bibinfo{pages}{58005} (\bibinfo{year}{2007}).

\bibitem[{\citenamefont{Lu et~al.}(2008)\citenamefont{Lu, Korniss, and
  Szymanski}}]{lu2008naming}
\bibinfo{author}{\bibfnamefont{Q.}~\bibnamefont{Lu}},
  \bibinfo{author}{\bibfnamefont{G.}~\bibnamefont{Korniss}}, \bibnamefont{and}
  \bibinfo{author}{\bibfnamefont{B.}~\bibnamefont{Szymanski}},
  \bibinfo{journal}{Phys. Rev. E} \textbf{\bibinfo{volume}{77}},
  \bibinfo{pages}{16111} (\bibinfo{year}{2008}).

\bibitem[{\citenamefont{Baronchelli}(2011)}]{baronchelli2011role}
\bibinfo{author}{\bibfnamefont{A.}~\bibnamefont{Baronchelli}},
  \bibinfo{journal}{Phys. Rev. E} \textbf{\bibinfo{volume}{83}},
  \bibinfo{pages}{046103} (\bibinfo{year}{2011}).

\bibitem[{\citenamefont{Caldarelli}(2007)}]{caldarelli2007sfn}
\bibinfo{author}{\bibfnamefont{G.}~\bibnamefont{Caldarelli}},
  \emph{\bibinfo{title}{{Scale-Free Networks: Complex Webs in Nature and
  Technology}}} (\bibinfo{publisher}{Oxford University Press},
  \bibinfo{address}{Oxford}, \bibinfo{year}{2007}).

\bibitem[{\citenamefont{Barrat et~al.}(2008)\citenamefont{Barrat,
  Barth\'{e}lemy, and Vespignani}}]{barratbook}
\bibinfo{author}{\bibfnamefont{A.}~\bibnamefont{Barrat}},
  \bibinfo{author}{\bibfnamefont{M.}~\bibnamefont{Barth\'{e}lemy}},
  \bibnamefont{and}
  \bibinfo{author}{\bibfnamefont{A.}~\bibnamefont{Vespignani}},
  \emph{\bibinfo{title}{Dynamical Processes on Complex Networks}}
  (\bibinfo{publisher}{Cambridge University Press},
  \bibinfo{address}{Cambridge}, \bibinfo{year}{2008}).

\bibitem[{\citenamefont{Dorogovtsev and Mendes}(2003)}]{mendesbook}
\bibinfo{author}{\bibfnamefont{S.~N.} \bibnamefont{Dorogovtsev}}
  \bibnamefont{and} \bibinfo{author}{\bibfnamefont{J.~F.~F.}
  \bibnamefont{Mendes}}, \emph{\bibinfo{title}{Evolution of networks: From
  biological nets to the {I}nternet and {WWW}}} (\bibinfo{publisher}{Oxford
  University Press}, \bibinfo{address}{Oxford}, \bibinfo{year}{2003}).

\bibitem[{\citenamefont{Pastor-Satorras and Vespignani}(2004)}]{romuvespibook}
\bibinfo{author}{\bibfnamefont{R.}~\bibnamefont{Pastor-Satorras}}
  \bibnamefont{and}
  \bibinfo{author}{\bibfnamefont{A.}~\bibnamefont{Vespignani}},
  \emph{\bibinfo{title}{Evolution and structure of the Internet: A statistical
  physics approach}} (\bibinfo{publisher}{Cambridge University Press},
  \bibinfo{address}{Cambridge}, \bibinfo{year}{2004}).

\bibitem[{\citenamefont{Boccaletti et~al.}(2006)\citenamefont{Boccaletti,
  Latora, Moreno, Chavez, and Hwang}}]{boccaletti2006cns}
\bibinfo{author}{\bibfnamefont{S.}~\bibnamefont{Boccaletti}},
  \bibinfo{author}{\bibfnamefont{V.}~\bibnamefont{Latora}},
  \bibinfo{author}{\bibfnamefont{Y.}~\bibnamefont{Moreno}},
  \bibinfo{author}{\bibfnamefont{M.}~\bibnamefont{Chavez}}, \bibnamefont{and}
  \bibinfo{author}{\bibfnamefont{D.}~\bibnamefont{Hwang}},
  \bibinfo{journal}{Phys. Rep.} \textbf{\bibinfo{volume}{424}},
  \bibinfo{pages}{175} (\bibinfo{year}{2006}).

\bibitem[{\citenamefont{Belykh et~al.}(2004)\citenamefont{Belykh, Belykh, and
  Hasler}}]{belykh2004connection}
\bibinfo{author}{\bibfnamefont{V.}~\bibnamefont{Belykh}},
  \bibinfo{author}{\bibfnamefont{I.}~\bibnamefont{Belykh}}, \bibnamefont{and}
  \bibinfo{author}{\bibfnamefont{M.}~\bibnamefont{Hasler}},
  \bibinfo{journal}{Physica D}
  \textbf{\bibinfo{volume}{195}}, \bibinfo{pages}{159} (\bibinfo{year}{2004}).

\bibitem[{\citenamefont{Frasca et~al.}(2008)\citenamefont{Frasca, Buscarino,
  Rizzo, Fortuna, and Boccaletti}}]{frasca2008synchronization}
\bibinfo{author}{\bibfnamefont{M.}~\bibnamefont{Frasca}},
  \bibinfo{author}{\bibfnamefont{A.}~\bibnamefont{Buscarino}},
  \bibinfo{author}{\bibfnamefont{A.}~\bibnamefont{Rizzo}},
  \bibinfo{author}{\bibfnamefont{L.}~\bibnamefont{Fortuna}}, \bibnamefont{and}
  \bibinfo{author}{\bibfnamefont{S.}~\bibnamefont{Boccaletti}},
  \bibinfo{journal}{Phys. Rev. Lett.} \textbf{\bibinfo{volume}{100}},
  \bibinfo{pages}{44102} (\bibinfo{year}{2008}).

\bibitem[{\citenamefont{Fujiwara et~al.}(2011)\citenamefont{Fujiwara, Kurths,
  and D{\'\i}az-Guilera}}]{albert2011sync}
\bibinfo{author}{\bibfnamefont{N.}~\bibnamefont{Fujiwara}},
  \bibinfo{author}{\bibfnamefont{J.}~\bibnamefont{Kurths}}, \bibnamefont{and}
  \bibinfo{author}{\bibfnamefont{A.}~\bibnamefont{D{\'\i}az-Guilera}},
  \bibinfo{journal}{Phys. Rev. E} \textbf{\bibinfo{volume}{83}},
  \bibinfo{pages}{025101} (\bibinfo{year}{2011}).

\bibitem[{\citenamefont{Wittgenstein}(1953)}]{wittgenstein53english}
\bibinfo{author}{\bibfnamefont{L.}~\bibnamefont{Wittgenstein}},
  \emph{\bibinfo{title}{Philosophical Investigations. (Translated by Anscombe,
  G.E.M.)}} (\bibinfo{publisher}{Basil Blackwell}, \bibinfo{address}{Oxford,
  UK}, \bibinfo{year}{1953}).

\bibitem[{\citenamefont{Steels}(1995)}]{Steels1996}
\bibinfo{author}{\bibfnamefont{L.}~\bibnamefont{Steels}},
  \bibinfo{journal}{Artificial Life} \textbf{\bibinfo{volume}{2}},
  \bibinfo{pages}{319} (\bibinfo{year}{1995}).

\bibitem[{\citenamefont{Angluin}(1980)}]{angluin1980global}
\bibinfo{author}{\bibfnamefont{D.}~\bibnamefont{Angluin}}, in
  \emph{\bibinfo{booktitle}{Proc. 12th Symposium on the Theory of Computing}}
  (\bibinfo{year}{1980}), pp. \bibinfo{pages}{82--93}.

\bibitem[{\citenamefont{Baronchelli
  et~al.}(2006{\natexlab{a}})\citenamefont{Baronchelli, Felici, Loreto,
  Caglioti, and Steels}}]{Baronchelli_JStatMech_2006}
\bibinfo{author}{\bibfnamefont{A.}~\bibnamefont{Baronchelli}},
  \bibinfo{author}{\bibfnamefont{M.}~\bibnamefont{Felici}},
  \bibinfo{author}{\bibfnamefont{V.}~\bibnamefont{Loreto}},
  \bibinfo{author}{\bibfnamefont{E.}~\bibnamefont{Caglioti}}, \bibnamefont{and}
  \bibinfo{author}{\bibfnamefont{L.}~\bibnamefont{Steels}},
  \bibinfo{journal}{J. Stat. Mech.} \textbf{\bibinfo{volume}{P06014}}
  (\bibinfo{year}{2006}{\natexlab{a}}).

\bibitem[{\citenamefont{Baronchelli et~al.}(2008)\citenamefont{Baronchelli,
  Loreto, and Steels}}]{Baronchelli_ng_long}
\bibinfo{author}{\bibfnamefont{A.}~\bibnamefont{Baronchelli}},
  \bibinfo{author}{\bibfnamefont{V.}~\bibnamefont{Loreto}}, \bibnamefont{and}
  \bibinfo{author}{\bibfnamefont{L.}~\bibnamefont{Steels}},
  \bibinfo{journal}{Int. J. Mod. Phys. C} \textbf{\bibinfo{volume}{19}},
  \bibinfo{pages}{785} (\bibinfo{year}{2008}).

\bibitem[{\citenamefont{De~Vylder and Tuyls}(2006)}]{de2006reach}
\bibinfo{author}{\bibfnamefont{B.}~\bibnamefont{De~Vylder}} \bibnamefont{and}
  \bibinfo{author}{\bibfnamefont{K.}~\bibnamefont{Tuyls}}, \bibinfo{journal}{J.
  Theor. Bio.} \textbf{\bibinfo{volume}{242}}, \bibinfo{pages}{818}
  (\bibinfo{year}{2006}).

\bibitem[{\citenamefont{Dall'Asta
  et~al.}(2006{\natexlab{a}})\citenamefont{Dall'Asta, Baronchelli, Barrat, and
  Loreto}}]{dallasta06b}
\bibinfo{author}{\bibfnamefont{L.}~\bibnamefont{Dall'Asta}},
  \bibinfo{author}{\bibfnamefont{A.}~\bibnamefont{Baronchelli}},
  \bibinfo{author}{\bibfnamefont{A.}~\bibnamefont{Barrat}}, \bibnamefont{and}
  \bibinfo{author}{\bibfnamefont{V.}~\bibnamefont{Loreto}},
  \bibinfo{journal}{Phys. Rev. E} \textbf{\bibinfo{volume}{74}},
  \bibinfo{eid}{036105} (pages~\bibinfo{numpages}{13})
  (\bibinfo{year}{2006}{\natexlab{a}}).

\bibitem[{\citenamefont{Dall and Christensen}(2002)}]{dall2002random}
\bibinfo{author}{\bibfnamefont{J.}~\bibnamefont{Dall}} \bibnamefont{and}
  \bibinfo{author}{\bibfnamefont{M.}~\bibnamefont{Christensen}},
  \bibinfo{journal}{Phys. Rev. E} \textbf{\bibinfo{volume}{66}},
  \bibinfo{pages}{016121} (\bibinfo{year}{2002}).

\bibitem[{\citenamefont{Dall'Asta and Baronchelli}(2006)}]{dallasta06c}
\bibinfo{author}{\bibfnamefont{L.}~\bibnamefont{Dall'Asta}} \bibnamefont{and}
  \bibinfo{author}{\bibfnamefont{A.}~\bibnamefont{Baronchelli}},
  \bibinfo{journal}{J. Phys. A}
  \textbf{\bibinfo{volume}{39}}, \bibinfo{pages}{14851} (\bibinfo{year}{2006}).

\bibitem[{\citenamefont{Baronchelli
  et~al.}(2006{\natexlab{b}})\citenamefont{Baronchelli, Dall'Asta, Barrat, and
  Loreto}}]{ng_lowdim}
\bibinfo{author}{\bibfnamefont{A.}~\bibnamefont{Baronchelli}},
  \bibinfo{author}{\bibfnamefont{L.}~\bibnamefont{Dall'Asta}},
  \bibinfo{author}{\bibfnamefont{A.}~\bibnamefont{Barrat}}, \bibnamefont{and}
  \bibinfo{author}{\bibfnamefont{V.}~\bibnamefont{Loreto}},
  \bibinfo{journal}{Phys. Rev. E} \textbf{\bibinfo{volume}{73}},
  \bibinfo{pages}{015102} (\bibinfo{year}{2006}{\natexlab{b}}).

\bibitem[{\citenamefont{Dall'Asta
  et~al.}(2006{\natexlab{b}})\citenamefont{Dall'Asta, Baronchelli, Barrat, and
  Loreto}}]{dallasta06}
\bibinfo{author}{\bibfnamefont{L.}~\bibnamefont{Dall'Asta}},
  \bibinfo{author}{\bibfnamefont{A.}~\bibnamefont{Baronchelli}},
  \bibinfo{author}{\bibfnamefont{A.}~\bibnamefont{Barrat}}, \bibnamefont{and}
  \bibinfo{author}{\bibfnamefont{V.}~\bibnamefont{Loreto}},
  \bibinfo{journal}{Europhys. Lett.} \textbf{\bibinfo{volume}{73}},
  \bibinfo{pages}{969} (\bibinfo{year}{2006}{\natexlab{b}}).

\end{thebibliography}

\end{document}